\pdfoutput=1
\documentclass[iop]{emulateapj}  %#@!
\usepackage{amsmath}
\usepackage{graphicx}
\usepackage{txfonts}
\usepackage{hyperref}
\usepackage{rotating}
\DeclareFontFamily{U}{euc}{}% I chose euc because the chart is called Euler cursive
\DeclareFontShape{U}{euc}{m}{n}{<-6>eurm5<6-8>eurm7<8->eurm10}{}%
\DeclareSymbolFont{AMSc}{U}{euc}{m}{n} % I chose AMSc because AMSa and AMSb are defined in the amsfonts-package
\DeclareMathSymbol{\umu}{\mathord}{AMSc}{"16}

\slugcomment{To be submitted to ApJ Letters}

\shorttitle{Optical imaging polarimetry of LkCa 15}
\shortauthors{Thalmann et al.}

\begin{document}

\title{Optical imaging polarimetry of the L\lowercase{k}C\lowercase{a}\ 15 protoplanetary disk with SPHERE ZIMPOL\altaffilmark{$\star$}} %$@! {\star}
    
   \author{C. Thalmann\altaffilmark{1},    %\inst{1}
          G.D. Mulders\altaffilmark{2},
          M. Janson\altaffilmark{3},
          J. Olofsson\altaffilmark{4,5,6},
          M. Benisty\altaffilmark{7,8},
          H. Avenhaus\altaffilmark{9},
          S.P. Quanz\altaffilmark{1}, %\\
          H.M. Schmid\altaffilmark{1},
          T. Henning\altaffilmark{4},
          E. Buenzli\altaffilmark{4},
          F. M\'enard\altaffilmark{10},
          J.C. Carson\altaffilmark{11,4},
          A. Garufi\altaffilmark{1},
          S. Messina\altaffilmark{12},
          C. Dominik\altaffilmark{13},
          J. Leisenring\altaffilmark{14},
          G. Chauvin\altaffilmark{15},
          M.R. Meyer\altaffilmark{1}
          }

%   \institute{Institute for Astronomy, ETH Zurich, Wolfgang-Pauli-Strasse 27, 
%   			8093 Zurich, Switzerland\\
%            \email{thalmann@phys.ethz.ch}
%            }

    \altaffiltext{$\star$}{Based on data collected at the European Southern 
        Observatory, Chile (ESO Programme 60.A-9358(A)).}
    \altaffiltext{1}{ETH Zurich, Institute for Astronomy, Wolfgang-Pauli-Strasse 27, 
   		8093 Zurich, Switzerland; \email{thalmann@phys.ethz.ch}}
    \altaffiltext{2}{Lunar and Planetary Laboratory, The University of Arizona, 
        Tucson, AZ 85721, USA}
    \altaffiltext{3}{Department of Astronomy, Stockholm University, 106 91, Stockholm, Sweden}
    \altaffiltext{4}{Max-Planck-Institut f\"ur Astronomie, K\"onigstuhl 17, 69117,
        Heidelberg, Germany}
    \altaffiltext{5}{Instituto de F\'isica y Astronom\'ia, Facultad de Ciencias, Universidad de Valpara\'iso, Av.\ Gran Breta\~na 1111, Playa Ancha, Valpara\'iso, Chile
}  
    \altaffiltext{6}{ICM nucleus on protoplanetary disks, Universidad de Valpara\'iso, Av. Gran Breta\~na 1111, Valpara\'iso, Chile}
    \altaffiltext{7}{Universit\'e Grenoble Alpes, IPAG, F-38000 Grenoble, France}
    \altaffiltext{8}{CNRS, IPAG, F-38000 Grenoble, France}
    \altaffiltext{9}{Departamento de Astronom\'ia, Universidad de Chile, 
        Casilla 36-D, Santiago, Chile}
    \altaffiltext{10}{UMI-FCA, CNRS/INSU, France (UMI 3386)}
    \altaffiltext{11}{Department of Physics \& Astronomy, College of Charleston, 
        66 George Street, Charleston, SC 29424, USA}
    \altaffiltext{12}{INAF -- Catania Astrophysical Observatory, via S.~Sofia 78,
        I-95123 Catania, Italy}
    \altaffiltext{13}{Anton Pannekoek Institute, University of Amsterdam, Science Park 904, 1098 XH Amsterdam, The Netherlands}
    \altaffiltext{14}{Steward Observatory, Department of Astronomy, University
    of Arizona, 933 N.\ Cherry Ave, Tucson, AZ 85721, USA}
    \altaffiltext{15}{LAOG, 414 Rue de la Piscine, Domaine Universitaire, BP 53,
    38041 Grenoble Cedex 09, France}
%   \date{Draft version}

% \abstract{}{}{}{}{} 
% 5 {} token are mandatory
 
  \begin{abstract}
      We present the first optical (590--890\,nm) 
      imaging polarimetry observations
      of the pre-transitional protoplanetary disk around the young solar analog
      LkCa~15, addressing a number of open questions raised by previous studies.
      We detect the previously unseen far side of the disk gap, confirm the 
      highly off-centered scattered-light gap shape that was postulated from 
      near-infrared imaging, at odds with the symmetric gap inferred from 
      millimeter interferometry.  
      Furthermore, we resolve the inner disk for the first time and trace it
      out to 30\,AU.  This new source of scattered light may contribute
      to the near-infrared interferometric signal attributed to the protoplanet
      candidate LkCa~15~b, which lies embedded in the outer regions of 
      the inner disk.  Finally, we present a new model for the system
      architecture of LkCa~15 that ties these new findings together. These
      observations were taken during science verification of SPHERE 
      ZIMPOL and demonstrate this facility's performance for faint
      guide stars under adverse observing conditions.
  \end{abstract}


%            -

   \maketitle
   
%
%________________________________________________________________

%\sloppy

%%%%%%%%%%%%%%%%%%%%%%%%%%%%%%%%%%%%%%%%%%%%%%%%%%%%%%%%%%%%%%%%%%%%%%%%
%%%%%%%%%%%%%%%%%%%%%%%%%%%%%%%%%%%%%%%%%%%%%%%%%%%%%%%%%%%%%%%%%%%%%%%%
%%%%%%%%%%%%%%%%%%%%%%%%%%%%%%%%%%%%%%%%%%%%%%%%%%%%%%%%%%%%%%%%%%%%%%%%
%%%%%%%%%%%%%%%%%%%%%%%%%%%%%%%%%%%%%%%%%%%%%%%%%%%%%%%%%%%%%%%%%%%%%%%%

\section{Introduction}

Transitional disks are %a class of objects with similar properties as 
protoplanetary disks %, but 
with heavily depleted gaps or cavities in their inner regions \citep[e.g.][]{strom1989,calvet2005}. %Because of this morphology, they 
They are thought to represent a transitional state in %the 
disk evolution, in which the gas-rich primordial disk gradually 
disperses %from the inside out %, either as a result of 
due to planet formation or %due to a variety of 
other effects \citep[e.g.][]{bryden1999,owen2012,alexander2014}.

One transitional disk target that has garnered particular attention in recent years is \object{LkCa~15}, %which is 
a 3--5 Myr old $\sim$1~$M_{\odot}$ star located approximately 140 pc away \citep{simon2000} in the Taurus-Auriga region. %The presence of 
A wide gap in the disk extending out to $\sim$50\,AU can be inferred from the infrared spectral energy distribution (SED; \citealt{espaillat2007, espaillat2008}), but it has also been spatially resolved both in scattered light at near-infrared wavelengths \citep{thalmann2010} and in thermal emission at mm/sub-mm wavelengths \citep{pietu2007,andrews2011,isella2014}. In %the 
near-infrared images, the outer edge of the gap appears off-center from the star, suggesting an eccentric gap \citep{thalmann2014}, although the mm/sub-mm images are consistent with a circular gap %gap centered on the location of the star
\citep{andrews2011,isella2014}. In addition, \citet{kraus2012} discovered a protoplanet candidate at $\sim$16--21\,AU using Sparse Aperture Masking (SAM). 
While some follow-up observations only provided upper limits
\citep{isella2014,whelan2015}, newest results from H$\alpha$ differential 
imaging confirm the accreting protoplanet hypothesis (K.~Follette, p.c.).

%If confirmed, a protoplanet still undergoing accretion would provide valuable constraints on the conditions and timescales of planet formation. 

Here we present new scattered-light images of the LkCa~15 disk taken 
at visible wavelengths 
%with SPHERE/ZIMPOL 
using Polarimetric Differential Imaging (PDI). The observations 
%aim to address the uncertainties discussed above by providing 
provide better spatial resolution and a more robust determination of the distribution of reflected light from the disk than previous data based on infrared Angular Differential Imaging (ADI), in which forward modeling is necessary to account for self-subtraction effects \citep{thalmann2011,milli2012,thalmann2013}.

%%%%%%%%%%%%%%%%%%%%%%%%%%%%%%%%%%%%%%%%%%%%%%%%%%%%%%%%%%%%%%%%%%%%%%%%
%%%%%%%%%%%%%%%%%%%%%%%%%%%%%%%%%%%%%%%%%%%%%%%%%%%%%%%%%%%%%%%%%%%%%%%%
%%%%%%%%%%%%%%%%%%%%%%%%%%%%%%%%%%%%%%%%%%%%%%%%%%%%%%%%%%%%%%%%%%%%%%%%
%%%%%%%%%%%%%%%%%%%%%%%%%%%%%%%%%%%%%%%%%%%%%%%%%%%%%%%%%%%%%%%%%%%%%%%%

\section{Observations and data reduction}
\label{s:obs}

LkCa 15 was observed on two nights during %as part of the 
Science Verification 
Time for the newly commissioned high-contrast imager SPHERE
(Spectro-Polarimetric High-contrast Exoplanet REsearch; \citealt{beuzit2008})
at the European Southern Observatory's Very Large Telescope facility.
Its visible-light instrument, ZIMPOL (Zurich Imaging POLarimeter, 
\citealt{thalmann2008,schmid2012}), was used %in conjunction 
with the extreme
adaptive optics system (SAXO; \citealt{fusco2006}) to obtain deep imaging
polarimetry in the `Very Broad Band' filter spanning the $R$ and $I$ 
bands (590--890 nm).  ZIMPOL offers extreme polarimetric contrast \citep{schmid2012} at a pixel scale of 3.6\,mas\,$\times$\,7.2\,mas and a
field of view of 3\farcs5\,$\times$\,3\farcs5.

The first observation, on 2015-02-02, comprised seven polarimetric 
cycles of four half-wave plate positions (Stokes $+Q$, $-Q$, $+U$, $-U$),
each %of which was 
exposed for 8 $\times$ 20 seconds at a time, for a total 
integration time of 75 minutes.  The second observation was executed in 
the same way on 2015-02-12.

The %frame 
exposure time had been chosen to saturate the %core of the 
star's point-spread function (PSF) to maximize sensitivity at larger
separations.  However, 
the performance of the adaptive optics was poor and highly variable 
throughout both observations, which rendered the star unsaturated in most
frames and yielded a full width at half maximum (FWHM) of $\sim$35\,mas
($\approx 2 \lambda/D$).  This can be ascribed to LkCa~15's apparent 
magnitude of $R \approx 12$\,mag at the faint limit of SAXO's guide star 
specifications, combined with unfavorable weather conditions and high 
airmass ($\sim$1.5).  Furthermore, a ``comet-tail'' aberration of the 
PSF core is observed in some frames, likely caused by malfunction of the 
atmospheric dispersion corrector (ADC).  
%A more detailed description of
%these effects is given in Appendix~\ref{s:psf}.

After correcting the raw ZIMPOL data for cosmetics and field distortion,
we applied the PDI data reduction method detailed in \citet{avenhaus2014a}.
It includes the following steps for reducing instrumental polarization:
(1) flux equalization of each simultaneous pair of raw images in order
to remove the polarization of the stellar PSF, (2) double-quotient 
calculation of Stokes $Q$ and $U$ polarized flux images for each 
polarimetric cycle, (3) empirical correction of polarimetric throughput
differences in Stokes $Q$ and $U$, and (4) empirical correction of angular
misalignment of the half-wave plate.

Lastly, we transform the Cartesian coordinate system of the Stokes
formalism $(Q,U)$ into a polar coordinate system $(Q_\phi,U_\phi)$
in which positive $Q_\phi$ polarization is defined as azimuthal with
respect to the star \citep{benisty2015}.
%.  The coordinate transform 
%$(Q,U) \mapsto (Q_\phi,U_\phi)$ is described in \citet{benisty2015}.
Under the single-scattering assumption, the scattering 
polarization should appear in the $Q_\phi$ image (typically as 
positive signal), whereas the $U_\phi$ image
should contain no scattered-light signal and can
be used %as a diagnostic for instrumental polarization and 
to estimate 
the noise level in the observation \citep{avenhaus2014a}.  While this
assumption does not strictly hold
for inclined, optically thick disks, as is the case for LkCa~15, the 
resulting deviations are small and automatically accounted for in the
error estimation \citep{avenhaus2014b}.

Since we find no significant differences between the two epochs, we 
combine both data sets into one final set of Stokes $(I,Q_\phi,U_\phi)$
images, which improves the quality of the results.  Since the data are
sensitivity-limited, we apply no frame selection. We convolve the 
polarization images with a circular aperture of 7 pixels (= 25\,mas) in
diameter to reduce shot noise.

%%%%%%%%%%%%%%%%%%%%%%%%%%%%%%%%%%%%%%%%%%%%%%%%%%%%%%%%%%%%%%%%%%%%%%%%
%%%%%%%%%%%%%%%%%%%%%%%%%%%%%%%%%%%%%%%%%%%%%%%%%%%%%%%%%%%%%%%%%%%%%%%%
%%%%%%%%%%%%%%%%%%%%%%%%%%%%%%%%%%%%%%%%%%%%%%%%%%%%%%%%%%%%%%%%%%%%%%%%
%%%%%%%%%%%%%%%%%%%%%%%%%%%%%%%%%%%%%%%%%%%%%%%%%%%%%%%%%%%%%%%%%%%%%%%%

\section{Results}
\label{s:results}

%%%%%%%%%%%%%%%%%%%%%%%%%%%%%%%%%%%%%%%%%%%%%%%%%%%%%%%%%%%%%%%%%%%%%%%%
%%%%%%%%%%%%%%%%%%%%%%%%%%%%%%%%%%%%%%%%%%%%%%%%%%%%%%%%%%%%%%%%%%%%%%%%

\subsection{Imagery}
\label{s:images}

\begin{figure}[tbp] %$#@!
\centering
\includegraphics[width=\linewidth]{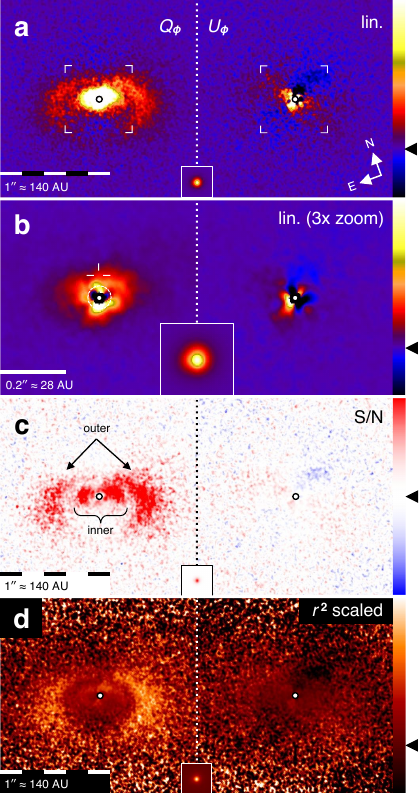}
%\begin{minipage}{0.49\linewidth} %$#@!
%\includegraphics[width=\linewidth]{Fig1_rev.pdf} %$#@!
%\end{minipage} %$#@!
%\begin{minipage}{0.49\linewidth} %$#@!
\vspace*{0.5mm}
\caption{\footnotesize Imaging polarimetry of LkCa~15 based on 
    SPHERE ZIMPOL observations at 590--890\,nm.  In each
    panel, the left and right sides show the $Q_\phi$ and $U_\phi$
    polarized flux images, respectively. Assuming single scattering,
    the disk appears as positive flux in the $Q_\phi$ 
    image, while the $U_\phi$ image %remains empty and thus 
    provides
    a noise estimate.  All images are convolved with a 7-pixel 
    (= 25\,mas) diameter circular aperture to reduce shot noise.
    The small insets show the shape of the PSF core.
    The location of the star is marked with black-and-white circles.
    The position angle of the North direction is $30^\circ$.  
    The black triangles indicate the zero level in the color scales.
    \textbf{(a)} Field of view of 1\farcs8 at a linear stretch with a
    maximum of $\sim${}$10^{-4}$ times  %7\times10^{-5}
    the brightest pixel of the stellar PSF (which is lightly saturated
    in some of the exposures). The known crescent of 
    scattered light from LkCa~15's outer disk is visible, along
    with a newly discovered inner disk component close to the
    star, saturating the color scale. 
    The white corners 
    delimit the area used for (b).
    \textbf{(b)} The central 0\farcs6 shown at 3x magnification at
    a linear stretch with a maximum of $\sim${}$10^{-3}$ times  
    %7\times10^{-4}
    the brightest pixel of the stellar PSF, showing an unsaturated 
    view of the inner disk component.
    The crosshairs indicate 
    the reported $K$-band location of LkCa~15~b in \citet{kraus2012}.
    The dashed circle marks the noise-dominated central area.
    \textbf{(c)} Signal-to-noise (S/N) maps at a stretch of $[-5,5]\sigma$.
    %The noise level as a function of separation is calculated as the 
    %standard deviation of the $U_\phi$ image in concentric annuli.  
    The
    same noise map is used for both the $Q_\phi$ and $U_\phi$ images.
    Both the outer disk and the new inner disk component are 
    detected at high significance.
    \textbf{(d)} The 1\farcs8 field of view with each pixel multiplied by 
    the square of its de-projected distance from the star, assuming an
    inclination of 50$^\circ$.
    }
\label{f:images}
%\end{minipage} %$#@!
\end{figure}

Our $RI$-band imaging polarimetry of LkCa~15 is shown in 
Fig.~\ref{f:images}.  Despite the bad PSF quality,
%of the individual exposures, 
the instantaneous nature of dual-beam polarimetry 
allowed us to remove the starlight effectively and resolve the LkCa~15
disk at high contrast and unprecedented angular resolution.  

The $Q_\phi$ image reveals %two prominent features: 
a crescent of polarized flux tracing the shape of the 
near-side edge of the outer disk seen in infrared angular 
ADI observations \citep{thalmann2010, thalmann2014}, and 
a previously unknown bright area of polarized flux surrounding the
star at small separations ($\lesssim 0\farcs25$, see panel 
\ref{f:images}b).  Both %of these
features appear consistently positive %as consistently positive signals
in the $Q_\phi$ image and are significantly brighter than the 
residuals in the $U_\phi$ image at similar radii, which identifies 
them as scattered light from circumstellar material.

Fig.~\ref{f:images}c provides
signal-to-noise (S/N) maps produced by dividing the pixel values of
star-centered annuli in the $Q_\phi$ and $U_\phi$ images by the 
standard deviation of the $U_\phi$ image in the same annuli.  Both 
features achieve S/N values above $5\sigma$ over a large
number of resolution elements and are therefore detected at high
confidence.  The inner disk is detected down to a separation of
$\sim$1\,FWHM (36 mas).

%To translate the scattered-light intensity into an estimate of
%the dust distribution
To estimate the dust distribution from the scattered-light intensity,
we compensate for the $r^{-2}$ fall-off of the illumination of
dust particles with physical distance $r$ from the star.  
We calculate $r$ for each pixel by deprojecting its
apparent distance from the star, assuming its light originates from
a plane inclined by 50$^\circ$ with its line of nodes at a position 
angle of 60$^\circ$ (cf.\ \citealt{thalmann2014}).  Fig.~\ref{f:images}d
shows the $Q_\phi$ and $U_\phi$ images after multiplying each pixel
value with $r^2$.  The disk gap is visible as an abrupt step in 
brightness at almost all position angles, including the faint far side
that had eluded all previous scattered-light observations.  The 
reduced brightness of the disk surface near the minor axis is likely
due to the dependence of scattering polarization on the phase angle.
However, we note that phase angles should change by at most
$\sim$6$^\circ$ along radial paths on the visible disk surface; thus,
the location of the gap edge should remain largely unaffected.

The eastern ansa appears brighter than the western one on the 
1\,$\sigma$ level (evaluated in the $r^2$-scaled image in 50-pixel 
diameter apertures).

% Cut away this paragraph to save space?
%We note that the residuals in the $U_\phi$ image %do not appear 
%completely random, but 
%include a faint quadrupole at the locus
%of the outer disk rim roughly antisymmetric about the %major and minor
%axes of the projected disk (Fig.~\ref{f:images}d). This is consistent 
%with the appearance of the inclined, optically thick disk of
%HD~100546 in \citet{avenhaus2014b}, and therefore likely represents
%a departure from strictly azimuthal polarization due to 
%second-order scattering effects.

%%%%%%%%%%%%%%%%%%%%%%%%%%%%%%%%%%%%%%%%%%%%%%%%%%%%%%%%%%%%%%%%%%%%%%%%
%%%%%%%%%%%%%%%%%%%%%%%%%%%%%%%%%%%%%%%%%%%%%%%%%%%%%%%%%%%%%%%%%%%%%%%%

\subsection{Geometric analysis}
\label{s:geo}

\begin{figure}[tbp] %[p]  %$#@!
\centering
\includegraphics[width=\linewidth]{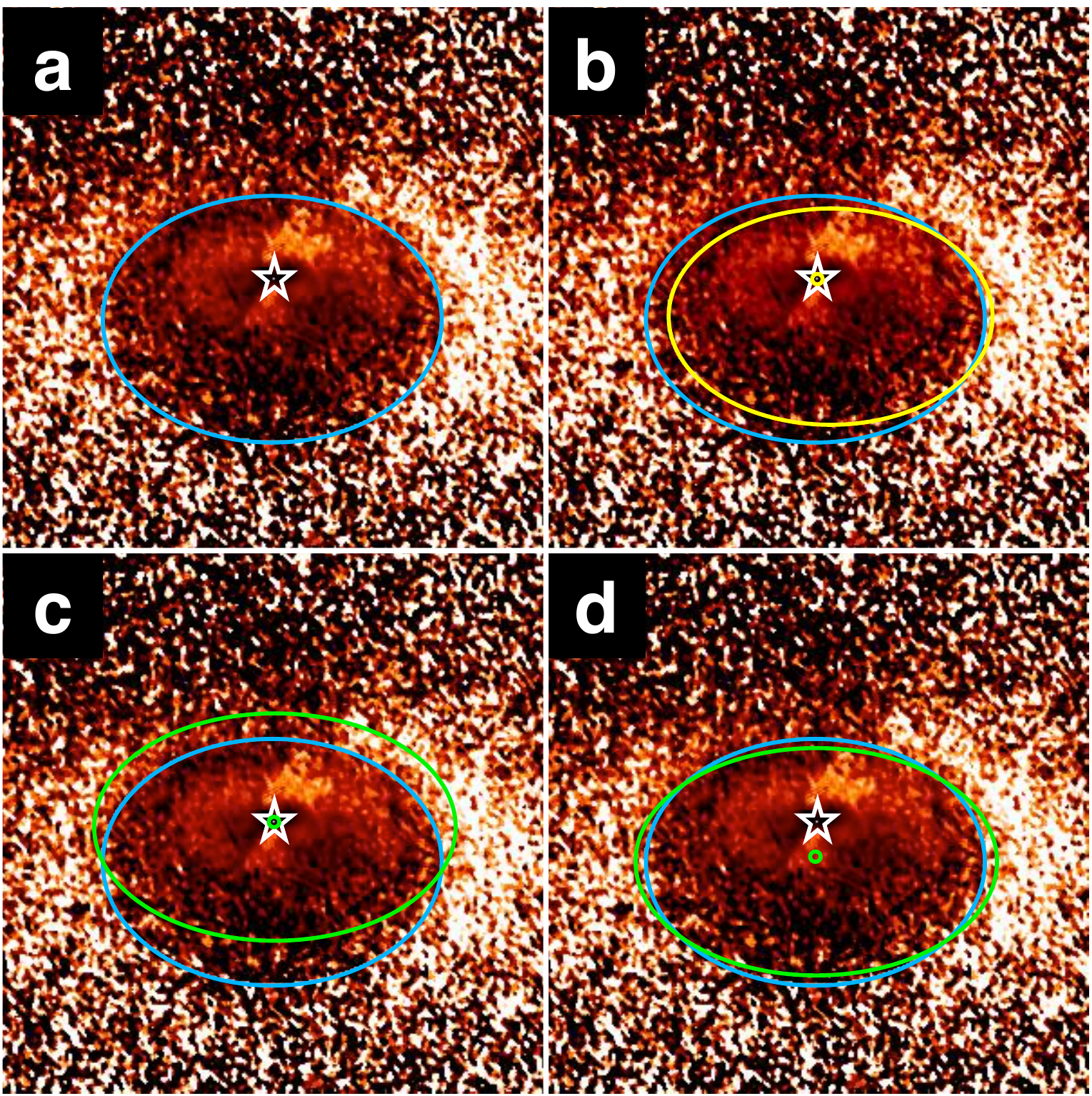} 
%\includegraphics[width=0.65\linewidth]{Fig2_rev.pdf} %$#@!
%\vspace*{1mm}
\caption{Ellipse fits to the LkCa~15 disk gap edge.  The 
    underlying image is the $r^2$-scaled $Q_\phi$ image as in
    Fig.~\ref{f:images}d displayed at a harder stretch, 
    whereas the light blue ellipse represents our
    fit to its gap (see text).  The location of the star is marked 
    with a star symbol.  
    \textbf{(a)} Visual confirmation of fit for the blue ellipse.
    \textbf{(b)} Modified maximum merit fit to the best-fit model
    from the ADI-based forward-modeling analysis in \citet{thalmann2014}
    overplotted in yellow.  The ellipse is registered against the 
    underlying image by its star position.
    \textbf{(c)} Fit ellipse (green) to the gap edge in 7\,mm 
    emission as 
    reported in Fig.~1 of \citet{isella2014}, registered by
    the assumed star position.
    \textbf{(d)} The same ellipse as in (c) aligned with our
    blue fit ellipse, demonstrating that the shapes match well. 
    The small green circle marks the true position of the star 
    relative to the green ellipse.
    }
\vspace*{2mm}
\label{f:ellipses}
\end{figure}

\begin{table}[tbp] %[p] %$#@!
\caption{Numerical results.}
\label{t:num}
\begin{tabular}{l|@{\quad}r@{\quad(}r@{, }r@{)\qquad }r@{\quad(}r@{, }r@{)}}
\hline
\multicolumn{7}{l}{\em{Fit ellipse to the disk gap edge in $Q_\phi$}}\\
\multicolumn{1}{l@{}}{} & \multicolumn{3}{@{}l@{}}{in mas} 
    & \multicolumn{3}{@{}l}{in proj.~AU}\\
Semimajor axis $a$        & 338 & $+11$   & $-18$
    & 47.4 & $+1.5$ & $-2.5$\\
Semiminor axis $b$        & 245 & $+21$   & $-11$
    & 34.3 & $+3.0$ & $-1.5$\\
Major-axis offset $x$     & $-3$ & $+11$  & $-14$
     & $-0.5$ & $+1.5$  & $-2.0$\\
Minor-axis offset $y$     & $-79$ & $+11$  & $-18$
    & $-11.1$ & $+1.5$  & $-2.5$\\[1mm]
\hline
\multicolumn{7}{l}{\em{Characteristic length scales of the inner scattering component}}\\
\multicolumn{1}{l@{}}{} & \multicolumn{3}{@{}l@{}}{in mas} 
   & \multicolumn{3}{@{}l}{in proj.~AU}\\
Major axis (East) $l_\mathrm{E}$ & 69 & $+15$ & $-11$
	& 9.7	& $+2.1$	& $-$$1.5$\\	
Major axis (West) $l_\mathrm{W}$ & 74 & $+11$ & $-9$
	& 10.4	& $+1.6$	& $-$$1.2$\\
Minor axis (North) $l_\mathrm{N}$ & 51 & $+6$ & $-5$
	& 7.2	& $+0.9$	& $-0.7$\\
Minor axis (South) $l_\mathrm{S}$ & 38 & $+10$ & $-6$
	& 5.3	& $+1.6$	& $-0.9$\\	[1mm]
\hline
\end{tabular}
\vspace*{2mm}
\end{table}

ADI observations of circumstellar disks suffer from flux loss and 
oversubtraction effects \citep{thalmann2011,milli2012,thalmann2013} and
therefore require forward modeling to interpret their results 
\citep{thalmann2014}.  Imaging polarimetry, on the other hand, yields
morphologically sound representations of the disk's appearance
in polarized light
\citep[e.g.,][]{hashimoto2011,quanz2013a,quanz2013b,garufi2013,benisty2015}.
% While uncertainties remain in
% reconstructing the disk's architecture from the polarized flux 
% distribution, sharp geometric disk features such as gaps and spirals
% can be readily identified

We characterize the 
%most striking geometric feature in our data---the
sharp gap edge in the $r^2$-scaled $Q_\phi$ image by fitting it with
an ellipse using a modified version of the maximum 
merit method introduced in \citet{thalmann2011}.  We generate a number
of ellipses with semimajor and -minor axes $a$, $b$ displaced from the
star's position by offsets $x$, $y$ along the two axes, respectively.
%To calculate the merit function of 
For a parameter set $(a,b,x,y)$, we 
consider the two concentric elliptical annuli bounded by the triplet
of ellipses $(a-\delta,b-\frac{b}{a}\delta,x,y)$, $(a,b,x,y)$, and
$(a+\delta,b+\frac{b}{a}\delta,x,y)$ with an annulus width of 
$\delta = 10$ pixels (= 36 mas = 1 FWHM). The merit function $\eta(a,b,x,y)$ 
is defined as the mean pixel value in the outer annulus of the
$r^2$-scaled $Q_\phi$ image minus the mean value of the inner annulus.  
The best fit is achieved when the merit function is maximized, i.e.,\
where the mean brightness increase from the inner to the outer annulus
is greatest.  As a measure of uncertainty, we define the ``well-fitting''
family of ellipses by a threshold of $\eta > \eta_\mathrm{max} - 
\sigma_\eta$, where $\sigma_\eta$ is the standard deviation of merits
of ellipses in the $U_\phi$ image. 
%, which is assumed to contain only noise.

As demonstrated in Fig.~\ref{f:ellipses}a, the best-fit ellipse matches
the visual locus of the gap edge convincingly.  Its numerical 
parameters and uncertainties are provided in Table~\ref{t:num}.
Figure~\ref{f:ellipses}b compares this fit to a second ellipse obtained
by applying the maximum merit method to the best-fit model from our
ADI forward-modeling analysis in \citet{thalmann2014}.  The two 
ellipses share similar sizes, aspect ratios, and significant minor-axis
offsets $y$. However, while we found a marginal major-axis offset $x$ 
towards the West in \citet{thalmann2014}, our present data are
consistent with $x=0$.

Fig.~\ref{f:ellipses}c shows the ellipse fit to the reconstructed 7\,mm
emission images reported in \citet{isella2014} superimposed on our
fit.  Their ellipse is centered on the star, 
which results in a significant mismatch with our 
off-centered ellipse.  While the two ellipses can be brought to 
coincide well with each other by translating one of them by
$y\approx80$\,mas (Fig.~\ref{f:ellipses}d), this offset is
highly significant compared to the astrometric accuracy of the
7\,mm interferometry (10\,mas, \citealt{isella2014}) and of the
ZIMPOL data ($<10$\,mas) and must therefore be physical.

%While disk gaps are known to vary in size as a 
%function of observed wavelength in systems with planet-induced dust 
%filtering \citep{dejuanovelar2013}, a lateral displacement of the gap 
%is more difficult to explain. 

%%%%%%%%%%%%%%%%%%%%%%%%%%%%%%%%%%%%%%%%%%%%%%%%%%%%%%%%%%%%%%%%%%%%%%%%
%%%%%%%%%%%%%%%%%%%%%%%%%%%%%%%%%%%%%%%%%%%%%%%%%%%%%%%%%%%%%%%%%%%%%%%%

\subsection{Surface brightness profiles}
\label{s:profiles}

\begin{figure}[tbp] %[p] %$#@!
\centering
\includegraphics[width=\linewidth]{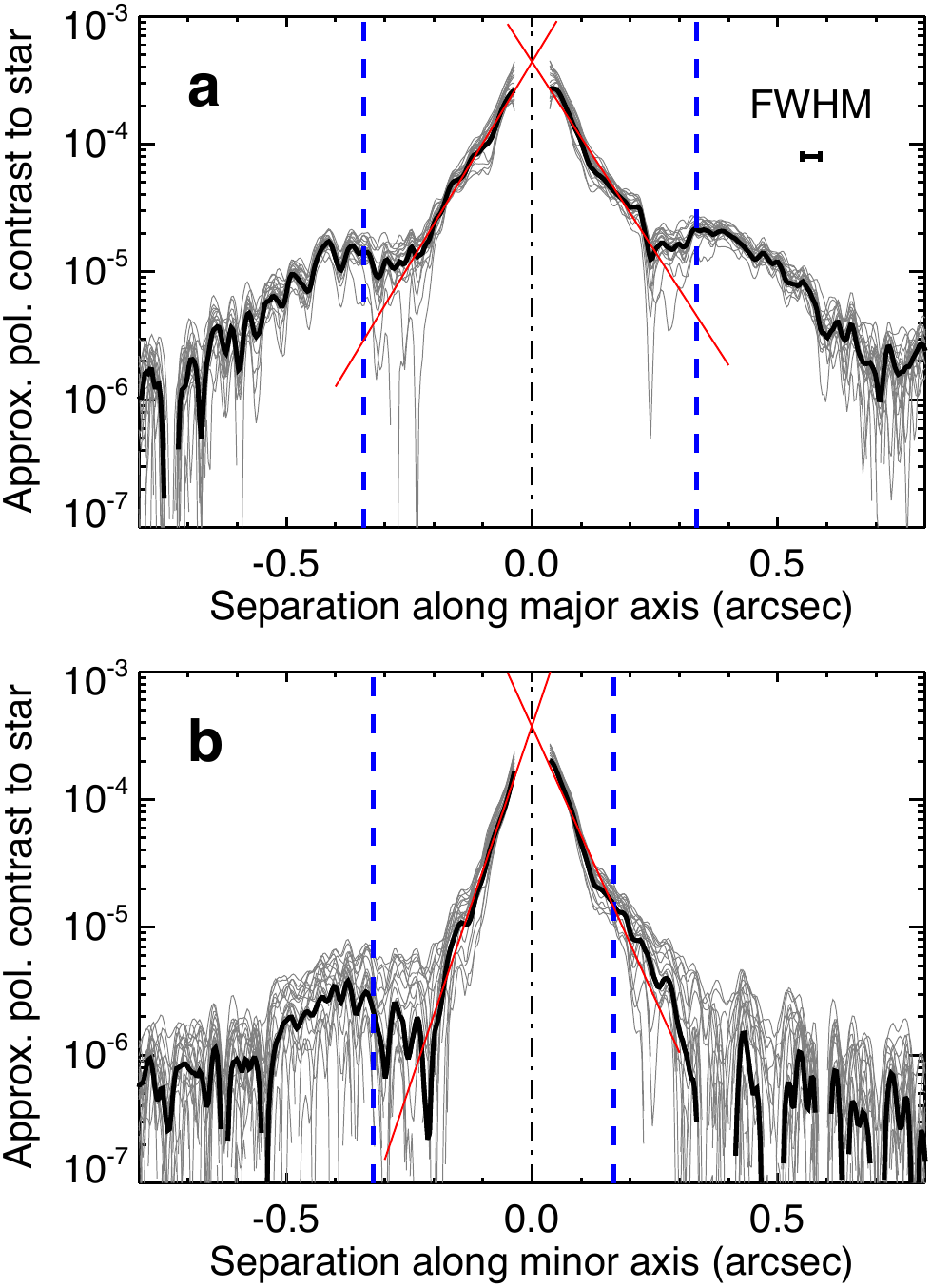}
\caption{Polarized surface brightness profiles for the LkCa~15    
    disk.
    \textbf{(a)} Profile of the $Q_\phi$ image along the major axis
    averaged over a width of 21 pixels (= 76 mas) centered on the star
    (thick black curve).
    The family of thin grey curves are obtained by adding similar 
    profiles taken in the $U_\phi$ image at various position angles to
    the black curve, and therefore visualize the noise level in the 
    data. The contrast of a 
    given pixel is calculated by the ratio of its value and the 
    maximum pixel in the stellar full-intensity PSF. 
    The normalization is approximate
    due to the stellar PSF being saturated in some exposures.  The blue
    dashed lines mark the location of the light blue fit ellipse in 
    Fig.~\ref{f:ellipses}. The red lines are exponential fits to the
    inner scattering component. Positive separation values correspond
    to the western side of the disk.
    The size of the PSF's FWHM is given for comparison.
    \textbf{(b)} Profile of the $Q_\phi$ image along the minor axis
    averaged over a width of 81 pixels (= 292 mas) centered on the 
    star. Positive separation values correspond to the northern side 
    of the disk.
    }
    \vspace*{1mm}
\label{f:profiles}
\end{figure}

%While the newly discovered  close to the star is statistically
%significant, its exact shape is difficult to determine.  In the following,
%we present surface brightness profiles as a first stage characterization.

Figures \ref{f:profiles}a and \ref{f:profiles}b plot the surface 
brightness profiles in the $Q_\phi$ image (without $r^2$ scaling) along 
the major and the minor axis, respectively.  The plotted values are 
pixel values as a function of separation averaged over a width
of 21 pixels (= 76\,mas) and 81 pixels (= 292\,mas), respectively, 
and normalized by the brightest pixel value of the star's intensity PSF.
Since some exposures in the data are saturated, the normalization is
only approximate.

The new inner disk component appears as a 
roughly exponential rise of surface brightness within a radius of
$\sim$0\farcs25.  We fit each side of the structure in each of the
two profiles as proportional to $\exp(-\theta/l)$, with $\theta$
the angular separation from the star and $l$ a characteristic 
length scale, using least-squares fitting.  These fits are listed 
in Table~\ref{t:num} and illustrated in Fig.~\ref{f:profiles}.

To gauge the uncertainty of these results, %, we again use the $U_\phi$
%as an estimate of the underlying noise in the $Q_\phi$ image.  
we 
extract a number of surface brightness profiles %as described above
from the noise-like $U_\phi$ image at various position angles, add 
each of them
to the science profiles from the $Q_\phi$ image, and recalculate the
exponential fits. The standard deviations of these ``disturbed'' $l$
values are then adopted as the error bars for the table values of $l$.
The disturbed surface brightness profiles are shown in grey in
Fig.~\ref{f:profiles}.

The polarized intensity is likely underestimated at 
the innermost separations due to overlapping PSF wings in the $Q$ and
$U$ images \citep{avenhaus2014b}.

%%%%%%%%%%%%%%%%%%%%%%%%%%%%%%%%%%%%%%%%%%%%%%%%%%%%%%%%%%%%%%%%%%%%%%%%
%%%%%%%%%%%%%%%%%%%%%%%%%%%%%%%%%%%%%%%%%%%%%%%%%%%%%%%%%%%%%%%%%%%%%%%%
%%%%%%%%%%%%%%%%%%%%%%%%%%%%%%%%%%%%%%%%%%%%%%%%%%%%%%%%%%%%%%%%%%%%%%%%
%%%%%%%%%%%%%%%%%%%%%%%%%%%%%%%%%%%%%%%%%%%%%%%%%%%%%%%%%%%%%%%%%%%%%%%%

\section{Discussion}
\label{s:discussion}

%%%%%%%%%%%%%%%%%%%%%%%%%%%%%%%%%%%%%%%%%%%%%%%%%%%%%%%%%%%%%%%%%%%%%%%%
%%%%%%%%%%%%%%%%%%%%%%%%%%%%%%%%%%%%%%%%%%%%%%%%%%%%%%%%%%%%%%%%%%%%%%%%

\subsection{Inner disk component}
\label{s:inner}

The newly discovered inner disk component, detected from $\sim$30\,AU 
down to the inner limit of $\sim$7\,AU (Fig.~\ref{f:profiles}), overlaps
with the location of the protoplanet candidate at $\sim$16--21\,AU 
derived from SAM observations
\citep{kraus2012}.  As with the outer disk, we expect anisotropic 
scattering to produce a distinct
%stark brightness contrast in full intensity
%between the near and far sides of the inner disk.  The 
full-intensity forward-scattering maximum on the near side, which
coincides with the position angle of the protoplanet candidate.  Taken
together with the forward-scattering maximum of the outer disk edge
%, which also passes through the field of view of the SAM observations
\citep{thalmann2014}, these disk features 
%provide a strong alternative
%explanation for the SAM closure phases that may obviate the need for a 
%protoplanetary point source, as in the case of T~Cha \citep{olofsson2013}.  
likely interfere with the reconstruction of SAM data, as in the case of
T~Cha \citep{olofsson2013}.
However, given the recent H$\alpha$ detection (K.~Follette, p.c.), a 
protoplanet
embedded in the inner disk similar to HD100546~b \citep{quanz2013a} 
remains the most plausible interpretation.

% and not connected to the observed gap structure 
% farther out at $\sim 50$ au.  % <== not sure what this means? --CTh
% ==> it means this is not the planet creating the disk wall, though I think that was never claimed in Kraus & Ireland. ---GDM

The inner disk component also sheds light on the extent of 
the gap as inferred from previous non-resolved observations. 
\Citet{espaillat2007} proposed a three-component structure 
% it said two-component earlier, but there are three components in 
% the following list...  we either have to call it three-component,
% or describe the optically thin component as a part of the inner
% disk...
% GDM: The espaillat model is a wall+optically thin dust. We could just call that an inner disk component
with an outer disk at $\sim$50\,AU consistent with the millimeter cavity, an 
inner disk starting at the dust sublimation radius to explain the 
near-infrared flux, and an optically thin component out to $\sim$5\,AU
to explain the prominent silicate emission feature. 
Our detection of scattered light between 7--30\,AU shows illuminated
dust is present at these locations above the disk mid plane.
%, likely tracing the surface of the gaseous disk.  % <== is this correct?
An extrapolation of the gap wall temperature 
($T\approx 100$\,K\,$/ \sqrt{a/a_\mathrm{wall}}$) % <== is this correct?
% this originally said T = 100 K /propto (...), which makes no sense...
% GDM: yeah that's fine, though it's proportional to 1/a. 
yields a temperature of 150--250\,K, consistent with the presence of
the 10 and 20\,$\mu$m silicate emission feature in the SED.  
We hypothesize that this new scattering structure represents 
%is part of an inner
%disk that is much larger than thought before, spanning 
the optically thin part of the inner disk extending out to at least 30\,AU. 
% GDM: let's not use 0.1 AU here, that number is not well constrained actually.
A quick adaptation of the disk model from 
\citet{mulders2010} indicates that this can be reconciled with 
the SED, since the contribution of the optically thin part to the SED is minor.
This geometry (narrow gap, depleted inner 
disk, millimeter cavity) qualitatively matches the 
imprint of a single planet on a disk \citep[e.g.][]{zhu2012,jangcondell2012}.
% <== is this the correct Zhu?  Please check...

%%%%%%%%%%%%%%%%%%%%%%%%%%%%%%%%%%%%%%%%%%%%%%%%%%%%%%%%%%%%%%%%%%%%%%%%
%%%%%%%%%%%%%%%%%%%%%%%%%%%%%%%%%%%%%%%%%%%%%%%%%%%%%%%%%%%%%%%%%%%%%%%%

\subsection{Off-centered gap}
\label{s:eccentric}

\begin{figure}[tbp] %[p] %$#@!
\centering
\includegraphics[width=\linewidth]{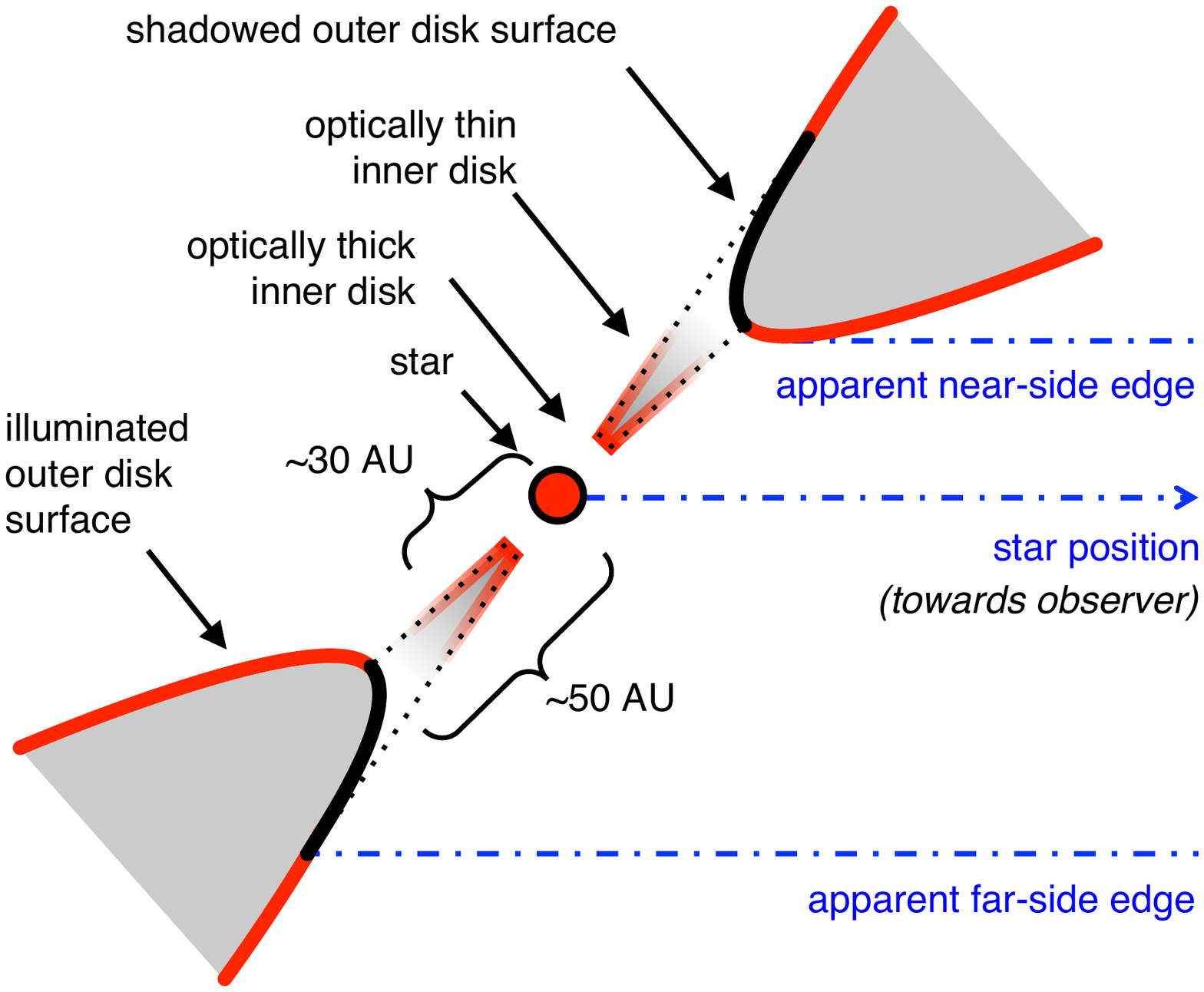}
\caption{Schematic of the proposed architecture of the 
    LkCa~15 disk (not to scale).  Red lining marks the disk
    surfaces illuminated by the star.  Dotted black lines 
    trace the edges of the shadow cast by the inner disk.
    }
\label{f:sketch}
\end{figure}

Our $RI$-band polarimetry confirms the prediction from $K$-band ADI
forward-modeling \citep{thalmann2014} that the outer boundary of the
gap in the LkCa~15 disk appears off-centered in reflected
light, with a highly significant offset of $y\approx80$\,mas (a third 
of the gap's projected radius) between the star and the gap's apparent 
center.  While disk gaps are known to vary in size as a function of 
observed wavelength in systems with planet-induced dust filtering 
\citep{dejuanovelar2013}, a lateral displacement of the gap is more 
difficult to justify and has not been predicted by numerical 
simulations.

The outer disk's vertical structure cannot account for this displacement 
since the gap wall profile is tapered (`round') rather than vertical 
\citep{thalmann2014}.  This is qualitatively confirmed by the wide,
homogeneous distribution of scattered light in Fig.~\ref{f:images}d.
% Too brief?

Another explanation is provided by shadowing. \Citet{espaillat2007}
originally proposed an inner disk coplanar with the outer disk, casting
a shadow onto the outer disk's inner wall \citet{mulders2010}.  % GDM: the espaillat paper didn't include the disk shadow actually
Our forward 
modeling in \citet{thalmann2014} achieved significantly better fits 
with an unshadowed gap wall, which led us to speculate that the inner
disk---and its shadow---were tilted out of the outer disk's plane.  On
the other hand, this arrangement should result in dark lanes cutting 
across the outer disk where the shadow plane intersects it, as seen in
HD~142527 \citep{marino2015}.  No such lanes are apparent in the 
LkCa~15 imagery.  

This conundrum can be resolved if the inner disk %plane approximately
%shares the outer disk's line of nodes, but 
has a slightly higher inclination than the outer disk
(Fig.~\ref{f:sketch}).
%This would cause the shadow to spill onto the upper surface of the 
%outer disk on the far side and onto the lower surface on the near
%side.  
The near-side disk rim as observed from Earth would then
be fully exposed to stellar irradiation, whereas the far-side 
disk surface would be obscured at the rim and only become visible
further out where the disk surface flares out of the shadow.  This 
would create an off-centered apparent gap in reflected light even with
a symmetric physical disk gap.  For the disk architecture in 
\citet{thalmann2014}, the requisite differential tilt would be
of order 1$^\circ$, which could be induced by a planet on
a tilted orbit (cf.\ \citealt{mouillet1997,augereau2001,lagrange2012}).
%While this scenario still implies that the gap is 
%larger in 7\,mm emission than in the optical, this matches 
%expectations from dust filtering \citep{dejuanovelar2013}.

A similar effect could be caused by a shadowing structure vertically
extending from an otherwise coplanar inner disk \citep{espaillat2011},
though this would imply dramatic variability on the timescale from 
days to years.  While we cannot exclude this possibility, we find
no significant variability in our observations 
(\citealt{thalmann2014}; this work).

%This scenario would likely require a planet to tilt the inner disk with
%respect to the outer 
%(cf.\ \citealt{mouillet1997,augereau2001,lagrange2012}); 
%thus, the observed eccentricity remains an indication of planets in the 
%LkCa~15 pre-transitional disk.

%%%%%%%%%%%%%%%%%%%%%%%%%%%%%%%%%%%%%%%%%%%%%%%%%%%%%%%%%%%%%%%%%%%%%%%%
%%%%%%%%%%%%%%%%%%%%%%%%%%%%%%%%%%%%%%%%%%%%%%%%%%%%%%%%%%%%%%%%%%%%%%%%

\begin{acknowledgements}
We thank A.~Isella and the anonymous referee for helpful correspondence.
This work has been carried out within the frame of the National Centre 
for Competence in Research PlanetS supported by the Swiss National 
Science Foundation. CT, SPQ, HMS, MRM acknowledge support from SNSF. 
CT is supported by the European Commission under Marie Curie IEF grant 
No.~329875.  JO acknowledges support from the Millennium Nucleus 
RC130007 (Chilean Ministry of Economy). JC was supported by NSF under 
award 1009203.
\end{acknowledgements}

\textit{Facilities:} \facility{VLT:Melipal (SPHERE ZIMPOL)}

%%%%%%%%%%%%%%%%%%%%%%%%%%%%%%%%%%%%%%%%%%%%%%%%%%%%%%%%%%%%%%%%%%%%%%%%
%%%%%%%%%%%%%%%%%%%%%%%%%%%%%%%%%%%%%%%%%%%%%%%%%%%%%%%%%%%%%%%%%%%%%%%%
%%%%%%%%%%%%%%%%%%%%%%%%%%%%%%%%%%%%%%%%%%%%%%%%%%%%%%%%%%%%%%%%%%%%%%%%
%%%%%%%%%%%%%%%%%%%%%%%%%%%%%%%%%%%%%%%%%%%%%%%%%%%%%%%%%%%%%%%%%%%%%%%%

%-------------------------------------------------------------------

\bibliographystyle{aa}

\begin{thebibliography}{}
\bibitem[Alexander et al.(2014)]{alexander2014} Alexander, R., 
Pascucci, I., Andrews, S., Armitage, P., 
\& Cieza, L.\ 2014, Protostars and Planets VI, 475 

\bibitem[Andrews et al.(2011)]{andrews2011} Andrews, S.~M., 
Rosenfeld, K.~A., Wilner, D.~J., \& Bremer, M.\ 2011, \apjl, 742, L5 

\bibitem[Augereau et 
al.(2001)]{augereau2001} Augereau, J.~C., Nelson, R.~P., Lagrange, A.~M., Papaloizou, J.~C.~B., \& Mouillet, D.\ 2001, \aap, 370, 447 

\bibitem[Avenhaus et al.(2014a)]{avenhaus2014a} Avenhaus, H., Quanz, 
S.~P., Schmid, H.~M., et al.\ 2014, \apj, 781, 87 

\bibitem[Avenhaus et al.(2014b)]{avenhaus2014b} Avenhaus, H., Quanz, 
S.~P., Meyer, M.~R., et al.\ 2014, \apj, 790, 56 

\bibitem[Benisty et al.(2015)]{benisty2015} Benisty, M., Juhasz, 
A., Boccaletti, A., et al.\ 2015, arXiv:1505.05325 

\bibitem[Beuzit et al.(2008)]{beuzit2008} Beuzit, J.-L., Feldt, 
M., Dohlen, K., et al.\ 2008, \procspie, 7014, 701418 

%\bibitem[Bouvier et 
%al.(2013)]{bouvier2013} Bouvier, J., Grankin, K., Ellerbroek, L.~E., Bouy, H., \& Barrado, D.\ 2013, \aap, 557, A77 

\bibitem[Bryden et al.(1999)]{bryden1999} Bryden, G., Chen, X., 
Lin, D.~N.~C., Nelson, R.~P., \& Papaloizou, J.~C.~B.\ 1999, \apj, 514, 344 

\bibitem[Calvet et al.(2005)]{calvet2005} Calvet, N., D'Alessio, 
P., Watson, D.~M., et al.\ 2005, \apjl, 630, L185 

\bibitem[de Juan Ovelar et 
al.(2013)]{dejuanovelar2013} de Juan Ovelar, M., Min, M., Dominik, C., et al.\ 2013, \aap, 560, A111 

\bibitem[Espaillat et al.(2007)]{espaillat2007} Espaillat, C., 
Calvet, N., D'Alessio, P., et al.\ 2007, \apjl, 670, L135 

\bibitem[Espaillat et al.(2008)]{espaillat2008} Espaillat, C., 
Calvet, N., Luhman, K.~L., Muzerolle, J., 
\& D'Alessio, P.\ 2008, \apjl, 682, L125 

\bibitem[Espaillat et al.(2011)]{espaillat2011} Espaillat, C., 
Furlan, E., D'Alessio, P., et al.\ 2011, \apj, 728, 49 

\bibitem[Fusco et al.(2006)]{fusco2006} Fusco, T., Rousset, G., 
Sauvage, J.-F., et al.\ 2006, Optics Express, 14, 7515 

\bibitem[Garufi et 
al.(2013)]{garufi2013} Garufi, A., Quanz, S.~P., Avenhaus, H., et al.\ 2013, \aap, 560, A105 

\bibitem[Hashimoto et al.(2011)]{hashimoto2011} Hashimoto, J., 
Tamura, M., Muto, T., et al.\ 2011, \apjl, 729, L17 

\bibitem[Isella et al.(2014)]{isella2014} Isella, A., Chandler, 
C.~J., Carpenter, J.~M., P{\'e}rez, L.~M., 
\& Ricci, L.\ 2014, \apj, 788, 129 

\bibitem[Jang-Condell 
\& Turner(2012)]{jangcondell2012} Jang-Condell, H., \& Turner, N.~J.\ 2012, \apj, 749, 153 

\bibitem[Kraus 
\& Ireland(2012)]{kraus2012} Kraus, A.~L., \& Ireland, M.~J.\ 2012, \apj, 745, 5 

\bibitem[Lagrange et 
al.(2012)]{lagrange2012} Lagrange, A.-M., Boccaletti, A., Milli, J., et al.\ 2012, \aap, 542, A40

\bibitem[Marino et al.(2015)]{marino2015} Marino, S., Perez, S., 
\& Casassus, S.\ 2015, \apjl, 798, L44 

\bibitem[Milli et 
al.(2012)]{milli2012} Milli, J., Mouillet, D., Lagrange, A.-M., et al.\ 2012, \aap, 545, A111 

\bibitem[Mouillet et al.(1997)]{mouillet1997} Mouillet, D., Larwood, 
J.~D., Papaloizou, J.~C.~B., \& Lagrange, A.~M.\ 1997, \mnras, 292, 896 

\bibitem[Mulders et 
al.(2010)]{mulders2010} Mulders, G.~D., Dominik, C., \& Min, M.\ 2010, \aap, 512, A11 

\bibitem[Olofsson et 
al.(2013)]{olofsson2013} Olofsson, J., Benisty, M., Le Bouquin, J.-B., et al.\ 2013, \aap, 552, A4 

\bibitem[Owen 
\& Jackson(2012)]{owen2012} Owen, J.~E., \& Jackson, A.~P.\ 2012, \mnras, 425, 2931 
\bibitem[Pi{\'e}tu et 
al.(2007)]{pietu2007} Pi{\'e}tu, V., Dutrey, A., \& Guilloteau, S.\ 2007, \aap, 467, 163 

%\bibitem[Rodigas et al.(2015)]{rodigas2015} Rodigas, T.~J., Stark, 
%C.~C., Weinberger, A., et al.\ 2015, \apj, 798, 96 

\bibitem[Quanz et al.(2013a)]{quanz2013a} Quanz, S.~P., Amara, A., 
Meyer, M.~R., et al.\ 2013, \apjl, 766, L1 

\bibitem[Quanz et al.(2013b)]{quanz2013b} Quanz, S.~P., Avenhaus, 
H., Buenzli, E., et al.\ 2013, \apjl, 766, L2 

\bibitem[Schmid et al.(2012)]{schmid2012} Schmid, H.-M., Downing, 
M., Roelfsema, R., et al.\ 2012, \procspie, 8446, 84468Y 

\bibitem[Simon et al.(2000)]{simon2000} Simon, M., Dutrey, A., 
\& Guilloteau, S.\ 2000, \apj, 545, 1034 

\bibitem[Strom et al.(1989)]{strom1989} Strom, K.~M., Strom, 
S.~E., Edwards, S., Cabrit, S., \& Skrutskie, M.~F.\ 1989, \aj, 97, 1451 

\bibitem[Thalmann et al.(2008)]{thalmann2008} Thalmann, C., Schmid, 
H.~M., Boccaletti, A., et al.\ 2008, \procspie, 7014, 70143F 

\bibitem[Thalmann et al.(2010)]{thalmann2010} Thalmann, C., Grady, 
C.~A., Goto, M., et al.\ 2010, \apjl, 718, L87 

\bibitem[Thalmann et al.(2011)]{thalmann2011} Thalmann, C., Janson, 
M., Buenzli, E., et al.\ 2011, \apjl, 743, L6 

\bibitem[Thalmann et al.(2013)]{thalmann2013} Thalmann, C., Janson, 
M., Buenzli, E., et al.\ 2013, \apjl, 763, L29 

\bibitem[Thalmann et 
al.(2014)]{thalmann2014} Thalmann, C., Mulders, G.~D., Hodapp, K., et al.\ 2014, \aap, 566, A51 

\bibitem[Whelan et al.(2015)]{whelan2015} Whelan, E.~T., Huelamo, 
N., Alcala, J.~M., et al.\ 2015, arXiv:1504.04824 

\bibitem[Zhu et al.(2012)]{zhu2012} Zhu, Z., Nelson, R.~P., 
Dong, R., Espaillat, C., \& Hartmann, L.\ 2012, \apj, 755, 6 


\end{thebibliography}

%\appendix
%
%\section{PSF behavior}
%\label{s:psf}

\clearpage

\end{document}